\documentclass[twocolumn,aps,preprintnumbers,superscriptaddress]{revtex4}
\usepackage{amsmath}
\usepackage{eurosym}
\usepackage{graphicx}
\usepackage{bm}
\usepackage{lipsum}
\usepackage{epsfig}
\usepackage{epstopdf}
\usepackage{colordvi}
\usepackage{float}
\usepackage{accents}
\usepackage{color}

\begin{document}

\title{A magic tilt angle for stabilizing two-dimensional solitons by
dipole-dipole interactions}
\author{Xing-You Chen}
\affiliation{Institute of Photonics Technologies, National Tsing Hua University, Hsinchu
30013, Taiwan}
\author{You-Lin Chuang}
\affiliation{Physics Division, National Center for Theoretical Sciences, Hsinchu 30013,
Taiwan}
\author{Chun-Yan Lin}
\affiliation{Institute of Photonics Technologies, National Tsing Hua University, Hsinchu
30013, Taiwan}
\author{Chien-Ming Wu}
\affiliation{Institute of Photonics Technologies, National Tsing Hua University, Hsinchu
30013, Taiwan}
\author{Yongyao Li}
\affiliation{School of Physics and Optoelectronic Engineering, Foshan University, Foshan 528000, China}
\author{Boris A. Malomed}
\affiliation{Department of Physical Electronics, School of Electrical Engineering,
Faculty of Engineering, and Center for Light-Matter Interaction,Tel Aviv
University, Tel Aviv 69978, Israel\\
ITMO University, St. Petersburg 197101, Russia}
\author{Ray-Kuang Lee}
\affiliation{Institute of Photonics Technologies, National Tsing Hua University, Hsinchu
30013, Taiwan}
\affiliation{Physics Division, National Center for Theoretical Sciences, Hsinchu 30013,
Taiwan}

\begin{abstract}
In the framework of the Gross-Pitaevskii equation, we study the formation
and stability of effectively two-dimensional solitons in dipolar
Bose-Einstein condensates (BECs), with dipole moments polarized at an
arbitrary angle $\theta$ relative to the direction normal to the system's
plane.
Using numerical methods and the variational approximation, we
demonstrate that unstable Townes solitons, created by the contact attractive
interaction, may be completely stabilized (with an anisotropic shape) by the
dipole-dipole interaction (DDI), in interval $\theta ^{\text{cr}}<\theta
\leq \pi /2$. The stability boundary, $\theta ^{\text{cr}}$, weakly depends
on the relative strength of DDI, remaining close to the
``magic angle", $\theta _{m}=\arccos \left( 1/\sqrt{3}\right) $. The results
suggest that DDIs provide a generic mechanism for the creation of stable
BEC\ solitons in higher dimensions.
\end{abstract}

\maketitle

%\pacs{42.65.Tg, 42.65.Sf, 42.70.Qs}
%\keywords{ Dipolar BEC solitons, nonlocal nonlinearity}

\section{Introduction}

The collisional interaction of matter waves in Bose-Einstein condensates
(BECs) resembles nonlinear interaction of optical waves in nonlinear
dielectric media~\cite{BEC}. If solely the attractive short-range $s$-wave
inter-atomic scattering is present in the BEC, which is tantamount to the
Kerr (cubic) nonlinearity in optics in the framework of the man-field
approximation, the two- and three-dimensional (2D and 3D)\ matter-wave
solitons are subject to the collapse-driven instability \cite{BAMalomed, Dum, Specialtopics}. In particular, the well-known instability of 2D Townes
solitons \cite{Townes} is induced by the critical collapse in the same
setting \cite{Sulem,Fibich}.

Long-range interactions may give rise to effects quite different from those
induced by the contact (local) cubic nonlinearity \cite{BEC-long}. In
particular, the experimental realization of BEC in gases of atoms carrying
large permanent magnetic moments ($\sim $several Bohr magnetons), \textit{viz%
}., $^{52}$Cr ~\cite{Pfau}, $^{164}$Dy~\citep{dy-gas}, and $^{168}$%
Er~\cite{er-gas}, has drawn a great deal of interest to effects of the
dipole-dipole interactions (DDIs), which are intrinsically anisotropic and
nonlocal \cite{Griesmaier,DDI-review}. Similar to the situation in nonlocal
optical media \cite{nonlocal, Assanto, Conti, WON}, the long-range nonlocal
nonlinearity may play a crucial role in the formation and stabilization of
solitons. A wide range of novel solitonic structures were predicted to be
supported by the nonlocal nonlinearities, such as discrete solitons~\cite%
{Yaroslav, mobile, Belgrade, Belgrade-2D}, azimuthons~\cite{LOPEZ}, solitary vortices
\cite{Briedis,CARMEL,Daniel}, vector solitons~\cite{YVK, AMG, lee},
dark-in-bright solitons \cite{Adhikari}, and other species of self-trapped
modes.

Even though trapping potentials can be used to stabilize 3D or quasi-2D
soliton condensates, dipolar BECs suffer from instabilities against
spontaneous excitation of roton and phonon modes at high and low momenta,
respectively \cite{d2d, maxon, bk, sr, phonon, sc}, which manifest
themselves at large strengths of DDI\ \cite{2D}. For matter waves trapped in
a cigar-shaped potential, existence of stable quasi-1D solitons was
predicted for combinations of the DDI and local interactions~\cite%
{1D-competing-dbec-09,soliton-molecule, TF-eq, am, Jiang, tunable-s,
2D-stability}. The DDI\ anisotropy brings the roton instability to trapped
dipolar gases in the 2D geometry, and drives the condensates into a
biconcave density distribution \cite{rotons-stability,roton-spectro}. Stable
strongly anisotropic quasi-2D solitons in the condensate with
in-plane-oriented dipolar moments have been predicted too \cite%
{anisotropic-soliton,Patrick}.

In this work, we consider a general setting for the formation of 2D bright
solitons supported by the contact interaction and DDI, with the dipoles
aligned at an arbitrary tilt angle with respect to the direction normal to
system's plane. By reducing the 3D Gross-Pitaevskii equation (GPE)\ to an
effective 2D equation for the ``pancake" geometry, we
establish conditions necessary for supporting matter-wave solitons in the
dipolar BEC, at different values of the DDI strength, chemical potential,
and tilt angle. In addition to the application of the well-known
Vakhitov-Kolokolov stability criterion~\cite{VK}, the linear-stability
analysis and variational approach are also used for the study of the
stability of the 2D dipolar soliton solutions. Starting with a fixed
strength of the attractive local interaction, our analysis reveals that the
originally unstable 2D Townes solitons may be stabilized with the help of
the DDI. It is thus found that 2D solitons are stable if the orientation
angle of the dipoles, with respect to the direction normal to the pancake's
plane, exceeds a certain critical (``magic") value, see Eq. (%
\ref{magic}) below, a similar ``magic angle" for the
sample's spinning axis being known in the theory of the nuclear magnetic
resonance ~\cite{magic-1, magic-2}. Thus, the DDI in dipolar gases provides
a generic mechanism for the soliton formation of stable 2D solitons.

The rest of the paper is structured as follows. In Sec. II, we outline the
derivation of the effective 2D model for the dipolar BEC polarized at an
arbitrary tilt angle, starting from the 3D Gross-Pitaevskii equation. Then,
in subsection II.A, numerical solutions for 2D solitons, based on this
effective equation, are produced for two different scenarios, which
correspond to small and large DDI strengths. In subsection II.B, a
variational solution is obtained by minimizing the corresponding Lagrangian,
using a 2D asymmetric Gaussian ansatz. The variational approximation (VA)
makes it also possible to predict the stability of the solitons on the basis
of the Vakhitov-Kolokolov (VK) criterion, which is an essential
result, as the stability is the critically important issue for the 2D
solitons. Further, in Sec. III, we display a stability map for 2D soliton
solutions in the parameter plane of the tilt angle and number of atoms,
produced by an accurate numerical solution of the stability-eigenvalue
problem for small perturbations. Comparison of the variational and numerical
results demonstrates that the VA predicts the ``magic
angle", as the stability boundary, quite accurately. In particular, while VA
produces the single value of the ``magic angle", given by
Eq. (\ref{magic}), which does not depend on the relative strength of the
DDI, $g_{d}$, with respect to the local self-attraction, the numerical
solution of the stability problem exhibits a very weak dependence on $g_{d}$%
. The paper is concluded by Sec. IV.

%As a consequence of Bochner's theorem, {\it i.e.}, every positive definite spectrum has a Fourier transform of non-negative Borel measure~\cite{Bochner},  it is proved that any nonlocal nonlinear response with a  positive definite spectrum  arrests  soliton collapse in arbitrary dimensions~\cite{o-bang2002}.

\begin{figure}[t]
\includegraphics[width=5.5cm]{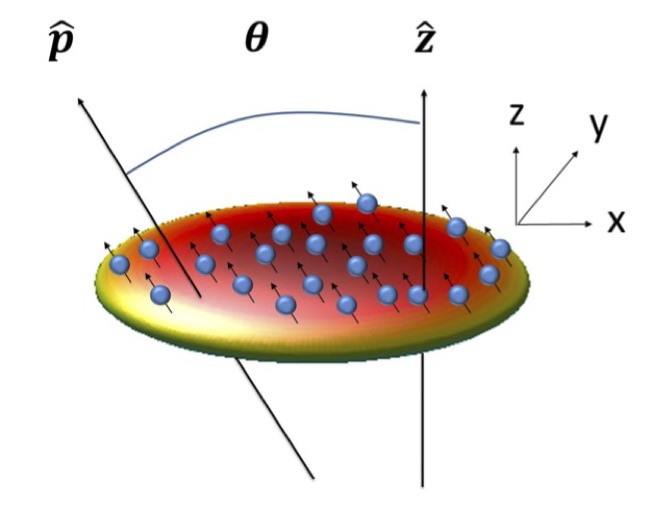}
\caption{A dipolar BEC in the 2D ``pancake" geometry, with
the dipole moments $\mathbf{\hat{p}}$ oriented along the tilt angle defined,
$\protect\theta $, with respect to the normal axis, $\mathbf{\hat{z}}$.}
\end{figure}

\section{The effective 2D model}

We consider an obliquely polarized dipolar BEC trapped in the pancake-shaped
potential, as shown in Fig. 1. The oblique orientation of dipole moments is
imposed by an external magnetic field, which makes the tilt angle, $\theta $%
, with the direction $\mathbf{\hat{z}}$ perpendicular to the pancake's
plane. The mean-field dynamics of the BEC at zero temperature is governed by
by the GPE, which includes the integral term accounting for the DDI \cite%
{DDI-review}:
\begin{eqnarray}
i\hbar \frac{\partial \Psi (\mathbf{r},t)}{\partial t}&=&\left[ -\frac{\hbar
^{2}}{2m}\nabla ^{2}+V(z)+g|\Psi (\mathbf{r},t)|^{2}\right.    \label{GPE} \\\notag
&+&\left. \left( \int d^{3}\mathbf{r}^{\prime }V_{d}(\mathbf{r}-\mathbf{r}%
^{\prime })|\Psi (\mathbf{r}^{\prime },t)|^{2}\right) \right] \Psi (\mathbf{r%
},t). 
\end{eqnarray}%
Here, $\Psi (\mathbf{r},t)$ is the wave function of condensate, $\mathbf{r}%
=(x,y,z)$ is the position vector, $m$ is the atomic mass, and $V(z)=m\omega
_{z}^{2}{z}^{2}/2$ is the confining potential acting in the transverse
direction. The anisotropic DDI kernel is
\begin{equation}
V_{d}(\mathbf{r})=g_{d}(1-3\cos ^{2}\eta )/r^{3},  \label{kernel}
\end{equation}%
where the DDI strength is $g_{d}=\mu _{0}\mu _{m}^{2}/4\pi $, with the
vacuum permeability $\mu _{0}$ and magnetic dipole moment $\mu _{m}$, while $%
\eta $ is the angle between vector $\mathbf{r}$ and the orientation of
dipole moments $\mathbf{\hat{p}}$. Note that this kernel vanishes at $\eta
=\arccos \left( 1/\sqrt{3}\right) $, which coincides with the
``magic angle" predicted by the VA as a boundary between
stable and unstable solitons, see Eq. (\ref{magic}) below. The usual contact
interaction is represented in Eq. (\ref{GPE}) by the local cubic term with
coefficient $g=4\pi \hbar ^{2}a/m$, where $a$ is the $s$-wave scattering
length $a$. The norm of the wave function is fixed by total number of atoms,
$N=\int d^{3}\mathbf{r}|\Psi (\mathbf{r},t)|^{2}$.

The 3D GPE (\ref{GPE}) can be reduced into an effective 2D equation,
provided that the confinement in the $z$-direction is strong enough. To this
end, we assume, as usual, that the 3D wave function is factorized, $\Psi (%
\mathbf{r})=\psi (\mathbf{\rho })\phi (z)\exp \left( -i\mu t/\hbar \right) $%
, with transverse coordinates $\mathbf{\rho }=(x,y)$ and chemical potential $%
\mu $ \cite{Luca, Proukakis, Luca2}. The transverse wave function is taken as the
normalized ground state of the respective trapping potential, $\phi
(z)=\left( \pi L_{z}^{2}\right) ^{-1/4}\exp \left( -z^{2}/2L_{z}^{2}\right) $%
, with the characteristic length $L_{z}=\sqrt{\hbar /m\omega _{z}}$. Then,
integrating Eq. (\ref{GPE}) over the $z$-coordinate, the factorized ansatz
leads one to the following effective 2D equation:
\begin{eqnarray}
&&\hspace{-0.2in}\left( \frac{\mu }{\hbar }-\frac{1}{2}\omega _{z}\right)
\hbar \psi ({\rho })=-\frac{\hbar ^{2}}{2m}\nabla _{\perp }^{2}\psi (\rho )+%
\frac{g}{\sqrt{2\pi }L_{z}}|\psi (\rho )|^{2}\psi (\rho )  \notag \\
&&\hspace{-0.2in}+\frac{g_{d}}{L_{z}}\left[ \int \frac{d^{2}\mathbf{k}_{\rho
}}{(2\pi )^{2}}\,n(\mathbf{k}_{\rho })V_{2d}\left( \frac{\mathbf{k}_{\rho
}L_{z}}{\sqrt{2}}\right) \,e^{i\,\mathbf{k}_{\rho }\cdot \rho }\right] \psi
(\rho ),  \label{2D-GPE}
\end{eqnarray}%
where $\nabla _{\perp }^{2}\equiv \partial ^{2}/\partial x^{2}+\partial
^{2}/\partial y^{2}$, $n(\mathbf{k}_{\rho })\equiv \int d^{2}\rho \,|\psi
(\rho )|^{2}\exp [-i\,\mathbf{k}_{\rho }\rho ]$ is the Fourier transform of
the 2D density, $|\psi (\rho )|^{2}$, and $k_{\rho
}=(k_{x}^{2}+k_{y}^{2})^{1/2}$. Further, defining that the dipoles are
polarized aligned in the $\left( x,z\right) $ plane, i.e., $\mathbf{\hat{p}}%
=(\sin \theta ,0,\cos \theta )$ and $\cos \eta =\mathbf{\hat{p}}\cdot \hat{%
\mathbf{r}}$, in the momentum ($\mathbf{k}$-) space, the DDI kernel takes
the form of
\begin{eqnarray}
&&V_{2d}\left( \frac{\mathbf{k}_{\rho }L_{z}}{\sqrt{2}}\right) =-\frac{2\sqrt{%
2\pi }}{3}(1-3\cos ^{2}\theta )  \\ \nonumber
&&+\left[ 1-3\cos ^{2}\theta +\cos (2\zeta )\sin ^{2}\theta \right] \pi
\mathbf{k}_{\rho }L_{z}\exp \left( \frac{\mathbf{k}_{\rho }^{2}L_{z}^{2}}{2}%
\right)\\\nonumber
&& \times \text{erfc}\left( \frac{\mathbf{k}_{\rho }L_{z}}{\sqrt{2}}\right) ,
\label{V2D}
\end{eqnarray}%
with $\cos \zeta \equiv k_{x}/k_{\rho }$ and the complementary error
function $\text{erfc}$ in the momentum space.

Rescaling Eq. (3) by $\mu \rightarrow \mu /\hbar \omega _{z}-1/2$%
, $\nabla _{\perp }\rightarrow \nabla _{\perp }L_{z}$, $\rho \rightarrow
\rho /L_{z}$, $\mathbf{k}_{\rho }\rightarrow \mathbf{k}_{\rho }L_{z}$, $\psi
(\rho )\rightarrow \psi (\rho )\sqrt{2(2\pi )^{1/2}|a|L_{z}}$, $%
g\rightarrow g/(4\pi \hbar \omega _{z} |a| L_{z}^2)$, and $%
g_{d}\rightarrow g_{d}/(2\sqrt{2\pi}\hbar \omega _{z} |a| L_{z}^2)$, we arrive at the
following normalized 2D equation:
\begin{eqnarray}
\mu \psi (\rho )&=&-\frac{1}{2}\nabla _{\perp }^{2}\psi (\rho )+g|\psi (\rho
)|^{2}\psi (\rho )  \label{scaled-2D}\\\notag
&+&g_{d}\left[ \int \frac{d^{2}\mathbf{k}_{\rho }}{(2\pi )^{2}}\,n(\mathbf{k}%
_{\rho })V_{2d}\left( \frac{\mathbf{k}_{\rho }}{\sqrt{2}}\right) \,e^{i\,%
\mathbf{k}_{\rho }\cdot \rho }\right] \psi (\rho ).
\end{eqnarray}%
According to the rescaling, the norm of the 2D wave function, $N_{2}\equiv
\int d^{2}\rho \,|\psi ({\rho })|^{2}$, is related to the number of atoms: $%
N=N_{2}\times (L_{z}/(2\sqrt{2\pi }|a|)$.

Our model is based on Eq. (\ref{scaled-2D}). For example, in the case of the
BECs of $^{52}$Cr atoms, the atomic magnetic moment is $\mu _{m}=6$ $\mu _{%
\mathrm{Bohr}}$, and an experimentally relevant trapping frequency,
\begin{equation}
\omega _{z}=2\pi \times 800~\text{Hz}  \label{Cr}
\end{equation}
\cite{Pfau-00, exp-Cr2, exp-Cr3, exp-Cr4}, corresponds to the characteristic
transverse length $L_{z}=0.493$ $\mathrm{\mu }$m.
With the same trapping frequency, for BECs of $^{168}$Er atoms, we have $\mu _{m}=7$ $\mu _{\mathrm{Bohr}}$ and $m=2.8\times 10^{-25}$ gram, which corresponds to a characteristic transverse length $L_{z}=0.274$  $\mathrm{\mu}$m; while for $^{162}$Dy atoms, we have $\mu _{m}=10$ $\mu _{\mathrm{Bohr}}$, $m=2.7\times 10^{-25}$ gram, and $L_{z}=0.279$  $\mathrm{\mu}$m, respectively.

\begin{figure}[t]
\includegraphics[width=8.4cm]{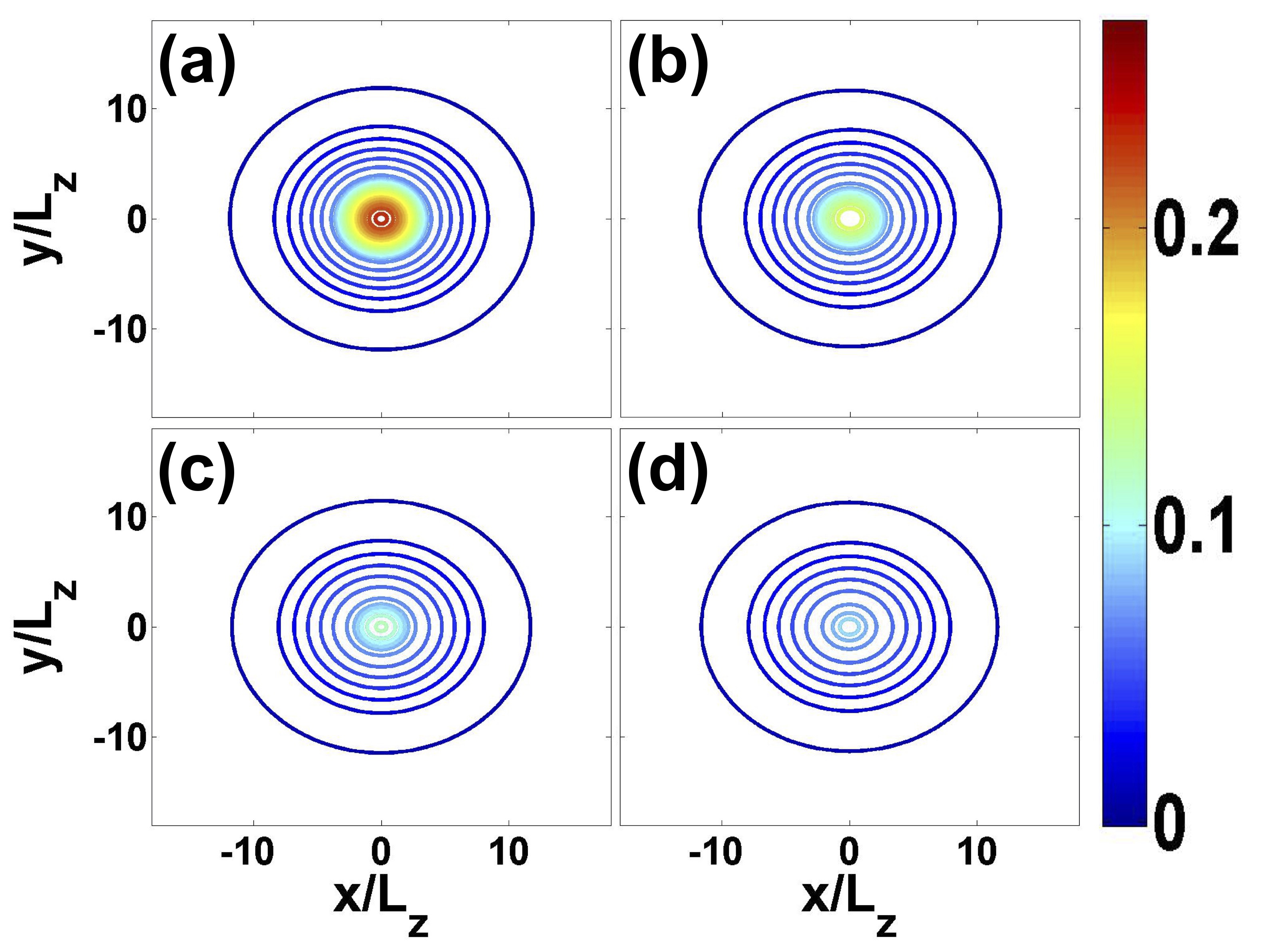}
\caption{Transverse profiles of 2D solitons in the $x$-$y$ plane normalized to the characteristic transverse length $L_{z}$, by the produced by the numerical
solution of Eq. (\protect\ref{scaled-2D}), with chemical potential $\protect%
\mu =-0.01$, fixed strength of contact attraction, $g=-1$, and a small DDI\
strength, $g_{d}=0.1$. The tilt angles are (a) $\protect\theta =0$, (b) $%
\protect\theta =0.73$ ($41.8^{\circ }$), (c) $\protect\theta =\protect\pi /3$%
, and (d) $\protect\theta =\protect\pi /2$.
Here, the corresponding particle number $N_2$ are (a) $7.15$, (b) $6.23$, (c) $5.71$ and (d) $5.34$, respectively.
}
\label{fig2}
\end{figure}
\begin{figure}[tbp]
\includegraphics[width=8.4cm]{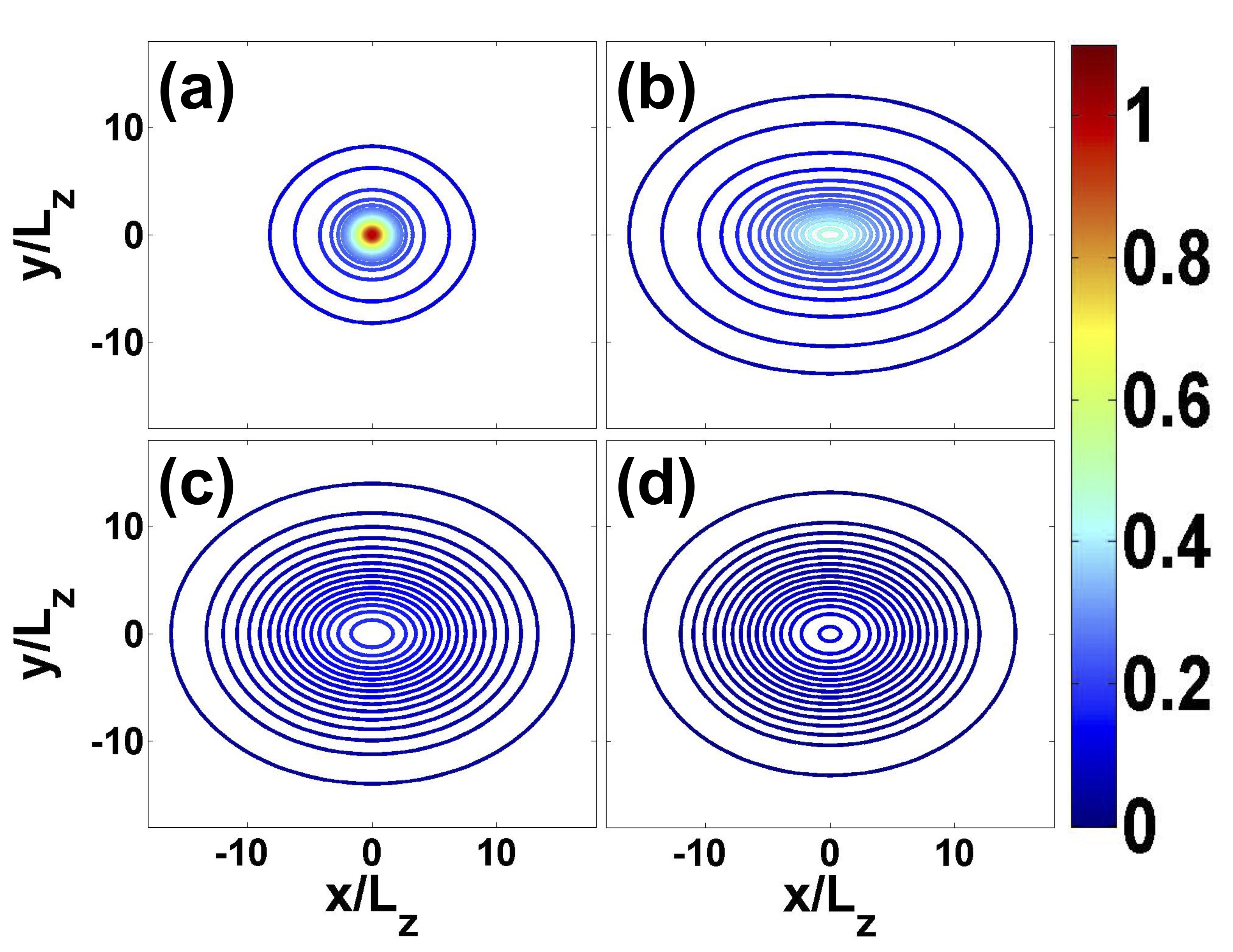}
\caption{The same as in Fig. \protect\ref{fig2}, but for stronger DDI, with $%
g_{d}=1.0$.Here, the corresponding particle number $N_2$ are (a) $4.69$, (b) $9.11$, (c) $4.53$, and (d) $2.86$, respectively.
}
\label{fig3}
\end{figure}

\subsection{2D numerical soliton solutions}

In the absence of the DDI, $g_{d}=0$, solutions in the form of isotropic
Townes solitons are supported by attractive contact interaction with $g<0$~%
\cite{Sulem,Fibich,soliton-book}. Then, by fixing the strength of the
contact attraction, $g=-1$, we introduce the DDI in Eq. (\ref{scaled-2D})
and seek for 2D bright-soliton solutions numerically, by varying the DDI
strength, $g_{d}$, for different values of of the chemical potential, $\mu $%
. The validity of our effective 2D equation for the pancake geometry is
ensured by checking that the transverse width of the 2D soliton solutions is
larger than the transverse-confinement length, $L_{z}$ in the $z$-direction.
This condition sets a constraint on the available range for the chemical
potential, i.e., $|\mu |/\hbar \omega _{z}\ll 1$. In our simulations, the 2D
effective equations remain valid in the range of%
\begin{equation}
-0.1<\mu <-0.003.  \label{range}
\end{equation}

The tilt angle of the dipoles in the $\left( x,z\right) $ plane was also
varied, in the full interval of $0<\theta <\pi /2$. The DDI sign is fixed as
$g_{d}>0$, which corresponds to the natural situation of the repulsion
between the dipoles oriented perpendicular to the pancake's plane, $\theta =0
$. Thus, the DDI is isotropic but repulsive at $\theta =0$, being
anisotropic at $\theta \neq 0$. Accordingly, the DDI tends to compete with
the fixed-strength contact attraction.

Numerical solution of Eq. (\ref{scaled-2D}) produces 2D soliton profiles,
typical examples of which are displayed in Figs. \ref{fig2} and \ref{fig3},
for $\mu =-0.01$. With the fixed contact-interaction coefficient, $g=-1$, we
find two different scenarios of the evolution of the shape of the 2D
solitons. For weak DDI, such as with coefficient $g_{d}=0.1$, starting with
the isotropic profile at $\theta =0$ [Fig. \ref{fig2}(a)], the transverse
widths in $x$- and $y$-directions both expand, but at different rates, as
the tilt angle increases, see Fig. \ref{fig2}(b-d) for $\theta =0.73$ ($%
41.8^{\circ }$) and $\pi /3$, respectively. The 2D solitons are wider along
the $x$-direction and narrower along $y$ because the dipoles are tilted in
the $\left( x,z\right) $ plane.

For a larger DDI\ strength, such as $g_{d}=1.0$, we still have an isotropic
profile at $\theta =0$, as shown in Fig. 3(a). As the tilt angle
increases, the transverse widths in $x$- and $y$-directions shrink just
slightly, remaining nearly equal at $\theta =0.73$ ($41.8^{\circ }$), $\pi /3
$, and $\pi /2$, as shown in Figs. 3(b-d).
Note that, quite naturally,
the radius of the isotropic profile, observed at $\theta =0$, is smaller in
Fig. \ref{fig2}(a) than in Fig. \ref{fig3}(a), as in the latter case the
dipole-dipole repulsion is much stronger than the competing contact
attraction. Nevertheless, the increase of $\theta $ makes the expansion of
the profiles and the growth of its anisotropy, which are effects of the DDI,
more salient in Fig. \ref{fig2}, i.e., when the DDI is weaker. This
counter-intuitive evolution of the shape may be explained by the fact that
it is shown not for the fixed number of atoms, $N_{2}$, but for a fixed
chemical potential, $\mu $. To keep the same $\mu $ in the case of the
stronger DDI competing with the contact self-attraction (in Fig. \ref{fig3}%
), the system needs to increase $N_{2}$, which, in turn, helps the contact
interaction to keep the compact nearly-isotropic shape of the soliton.

\begin{figure}[t]
\includegraphics[width=8.4cm]{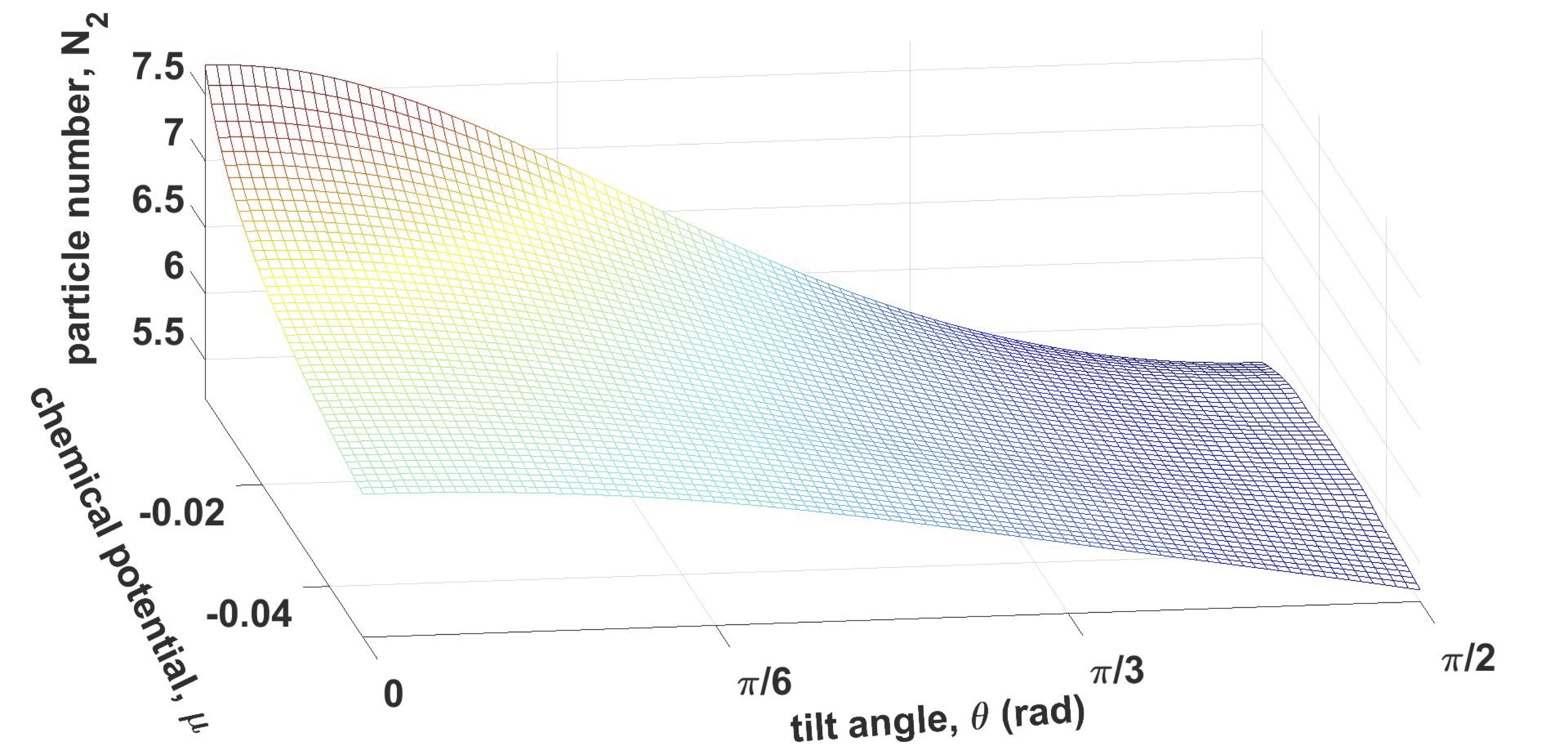}
\caption{The scaled 2D particle number, $N_{2}$ in the soliton
solutions,versus the tilt angle, $\protect\theta $, and chemical potential, $%
\protect\mu $, at the fixed strength of the contact interaction, $g=-1$, and
a small DDI strength, $g_{d}=0.1$.}
\label{fig4}
\end{figure}

\begin{figure}[tbp]
\includegraphics[width=8.4cm]{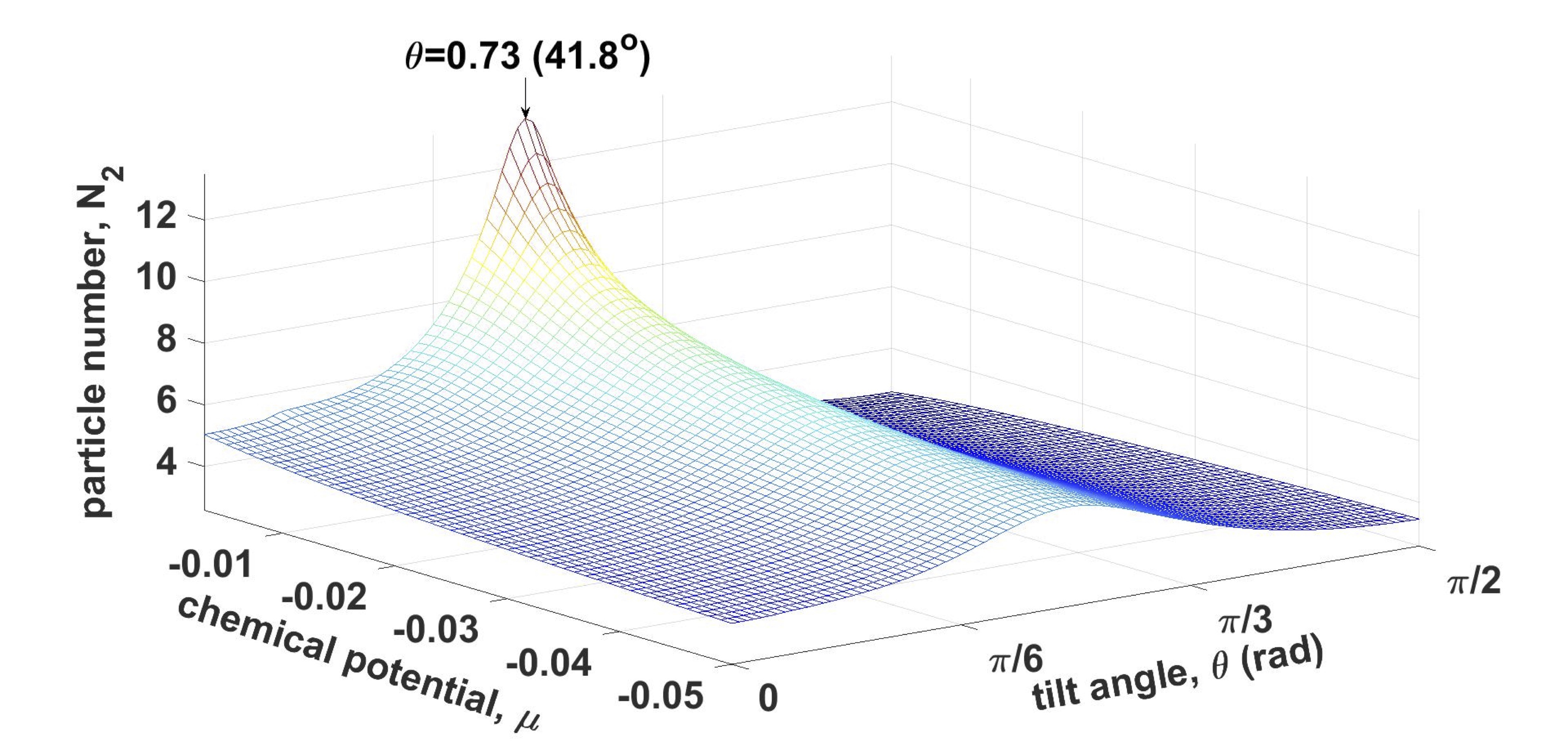} .
\caption{The same as in Fig. \protect\ref{fig4}, but for a much stronger
DDI, with $g_{d}=1.0$. In this case, $N_{2}$ attains ist maximum at $\protect%
\theta \approx 0.73$ ($41.8^{\circ }$), irrespective of the value of $%
\protect\mu $.}
\label{fig5}
\end{figure}

\begin{figure}[tbp]
\includegraphics[width=8.4cm, height=12.0cm]{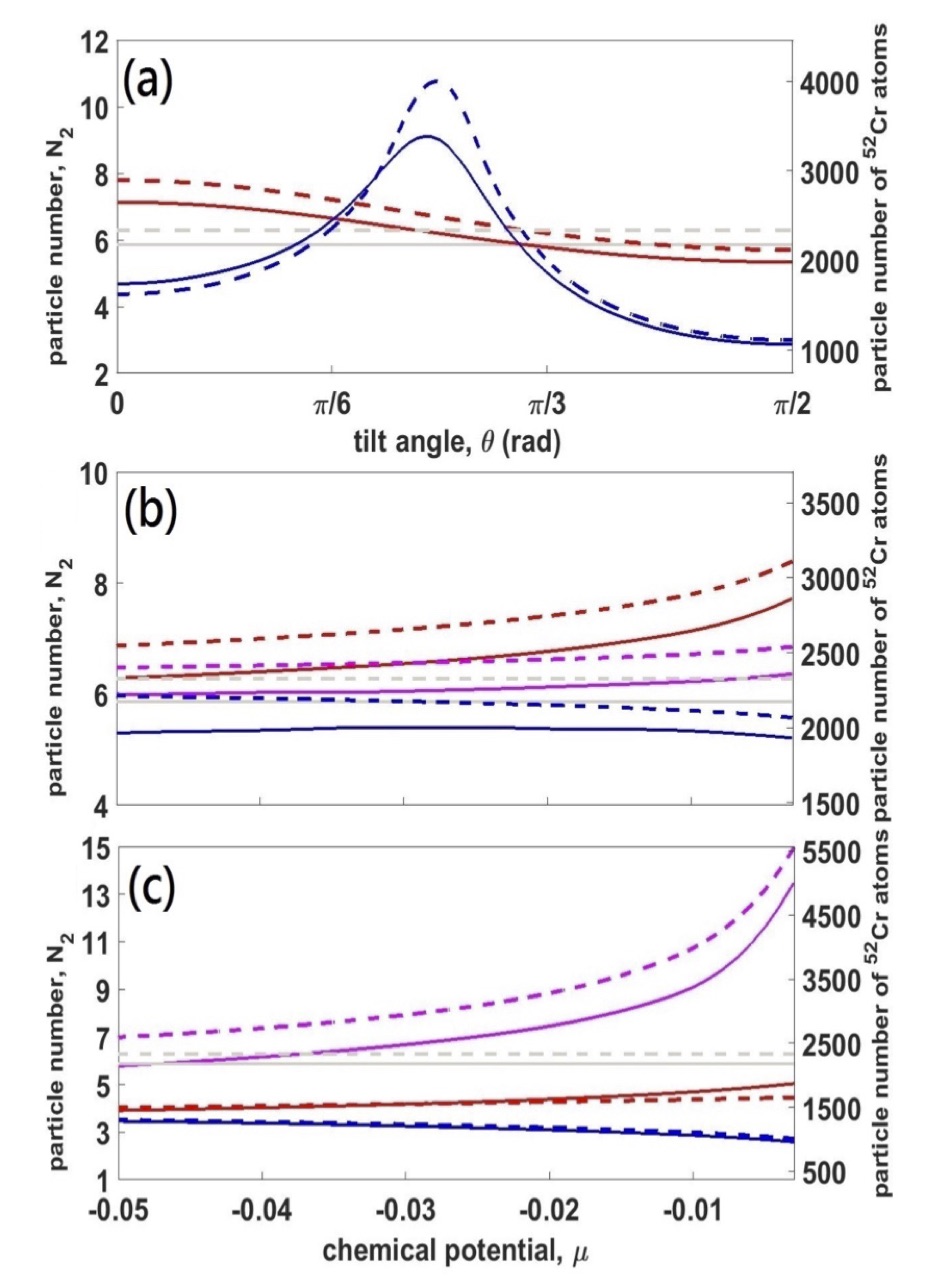}
\caption{
Comparison of the norm of the 2D wave function, $N_{2}$, as
produced by tne numerical solution and variational approximation (solid and
dashed lines, respectively). (a) $N_{2}(\protect\theta )$ at fixed $\protect%
\mu =-0.01$, for the weak and strong DDI, $g_{d}=0.1$ and $1$ (red and blue
lines, respectively). Two other panels display $N_{2}(\protect\mu )$ for the
same fixed values of the DDI strength: $g_{d}=0.1$ (b) and $1.0$ (c). In
these panels, fixed values of the tilt angle are $\protect\theta =0$ (red
lines), $\protect\theta = \protect\theta _{0}\approx 0.73$ [see Eq. (8); magenta], and $\protect\theta =\protect\pi /2$ (blue). In each
panel we also show, by solid and dashed black horizontal lines, the constant
value of $N_{2}$ for the Townes soliton (when the DDI is absent, $g_{d}=0$)
and its variationally predicted counterpart (see the main text for details).
For the condensate of $^{52}$Cr atoms, the corresponding total numbers of
atoms, for fixed values of other parameters [see Eq. (\protect\ref{Cr})],
are given on the right vertical axes, as a reference for a possible
experiment.}
\label{fig6}
\end{figure}

To present a clearer illustration on these trends, we display, in Figs. \ref%
{fig4} and \ref{fig5}, $N_{2}$ as a function of $\theta $ and $\mu $, for
the same small and large strengths of DDI, i.e., $g_{d}=0.1$ and $g_{d}=1.0$%
, respectively. In accordance with what is said above, $N_{2}$ decreases
monotonously at $g_{d}=0.1$, as the tilt angle increases from $\theta =0$ to
$\pi /2$, at all values of $\mu $. However, the stronger DDI strength (with $%
g_{d}=1.0$) produces a completely different picture (also in agreement with
the above explanation): as $\theta $ increases from $0$, $N_{2}$ at first
increases too, reaching a maximum at%
\begin{equation}
\theta =\theta _{0}\approx 0.73~(\text{equivalent to }41.8^{\circ })
\label{m}
\end{equation}%
[note that this angle is smaller than the critical
(``magic") one, $\theta _{m}$, given below by Eq. (\ref{magic}), which is an
approximate boundary between the stable and unstable solitons]. As mentioned
above, the increase of $N_{2}$ is necessary to keep the same value of $\mu $
while the essentially repulsive DDI competes with the local self-attraction,
at $\theta <\theta _{0}$. Then, $N_{2}$ decreases, as $\theta $ passes $%
\theta _{0}$ and approaches $\pi /2$. Indeed, in the latter case, the DDI
becomes essentially attractive \cite{anisotropic-soliton}, hence the local
and nonlocal interaction act together, instead of competing, making it possible
to keep the given value of $\mu $ with a smaller norm. Note that these
trends are the same at different values of $\mu $, although the
corresponding values of $N_{2}$ are, naturally, different. Below, we
demonstrate that angle $\theta _{0}$ can be accurately predicted by the
variational approximation, see Fig. \ref{fig6}(a).

\subsection{The variational approximation (VA)}

In addition to numerical solutions, we have developed the VA, following the
lines of Refs.~\cite{variation-1, variation} and using the Lagrangian
density corresponds to Eq. (\ref{scaled-2D})
\begin{eqnarray}
\mathcal{L}&=&-\mu |\psi |^{2}+\frac{1}{2}|\nabla _{\perp }\psi |^{2}+\frac{g}{%
2}|\psi |^{4}  \nonumber \\
&+&\frac{g_{d}}{2}|\psi |^{2}\int d^{2}\rho ^{\prime }\,V_{2d}\left( \rho
-\rho ^{\prime }\right) |\psi (\rho ^{\prime })|^{2}.  \label{L}
\end{eqnarray}%
The corresponding Gaussian ansatz is, naturally, anisotropic:
\begin{equation}
\psi _{\text{ans}}=\sqrt{\frac{N_{2}}{\pi w_{x}w_{y}}}\exp \left( -\frac{%
x^{2}}{2w_{x}^{2}}-\frac{y^{2}}{2w_{y}^{2}}\right) \,,
\end{equation}%
with the 2D norm $N_{2}$, and different transverse widths in $x$- and $y$%
-directions, $w_{x}$ and $w_{y}$.
Then, the
effective Lagrangian $L=\int dxdy\,\mathcal{L}$ is calculated:
\begin{eqnarray}
L&=&-N_{2}\mu +\frac{N_{2}(w_{x}^{2}+w_{y}^{2})}{4w_{x}^{2}w_{y}^{2}}  \notag
\\
&+&\frac{gN_{2}^{2}}{4\pi w_{x}w_{y}}+\frac{g_{d}N_{2}^{2}}{8\pi ^{2}}\left(
f_{1}+f_{2}\right) ,  \label{Leff}
\end{eqnarray}%
were we have introduced the short-hand notation:
\begin{eqnarray}
&&\hspace{-0.2in}f_{1}=-\frac{4\sqrt{2}\pi ^{3/2}}{3w_{x}w_{y}}\left(
1-3\cos ^{2}\theta \right),  \notag \\
&&\hspace{-0.2in}f_{2}=2\pi ^{2}\int_{0}^{\infty }dk_{\rho }\,\left\{
k_{\rho }^{2}\exp \left( \frac{2k_{\rho }^{2}-k_{\rho
}^{2}(w_{x}^{2}+w_{y}^{2})}{4}\right) \right. \notag  \\
&&\hspace{-0.2in} \times \text{erfc}\left( \frac{k_{\rho }}{\sqrt{2}}\right) \times \left[ (1-3\cos ^{2}\theta )\,I_{0}\left( \frac{k_{\rho }^{2}(w_{x}^{2}-w_{y}^{2})}{4}\right) \right. \notag \\
&&\hspace{-0.2in}\left.\left.  -\left( \sin ^{2}\theta \right) \,I_{1}\left( \frac{k_{\rho }^{2}(w_{x}^{2}-w_{y}^{2})}{4}\right) \right] \right\},
\label{f}
\end{eqnarray}%
with the modified Bessel functions, $I_{0,1}\left( z\right) $. The
Euler-Lagrange equations follow from Eq. (\ref{Leff}) in the form of $%
\partial L/\partial \left( w_{x, y},N_{2}\right) =0$:
\begin{gather}
\hspace{-0.3in}-\mu +\frac{w_{x}^{2}+w_{y}^{2}}{4w_{x}^{2}w_{y}^{2}}+\frac{%
gN_{2}}{2\pi w_{x}w_{y}}+\frac{g_{d}N_{2}(f_{1}+f_{2})}{4\pi ^{2}}=0,
\label{N} \\
\hspace{-0.3in}-\frac{1}{2w_{x}^{3}}-\frac{gN_{2}}{4\pi w_{x}^{2}w_{y}}+%
\frac{g_{d}N_{2}}{8\pi ^{2}}\left( \frac{\partial {f_{1}}}{\partial {w_{x}}}+%
\frac{\partial {f_{2}}}{\partial {w_{x}}}\right) =0,  \label{1} \\
\hspace{-0.3in}-\frac{1}{2w_{y}^{3}}-\frac{gN_{2}}{4\pi w_{x}w_{y}^{2}}+%
\frac{g_{d}N_{2}}{8\pi ^{2}}\left( \frac{\partial {f_{1}}}{\partial {w_{y}}}+%
\frac{\partial {f_{2}}}{\partial {w_{y}}}\right) =0.  \label{2}
\end{gather}%

For small arguments, $0<|z|\ll \sqrt{\alpha +1}$, the modified Bessel
function can be replaced by the first term of its expansion, $I_{\alpha
}(z)\approx (z/2)^{\alpha }/\Gamma (\alpha +1)$, where $\Gamma $ is the
Gamma-function. Such an approximation makes it possible to simplify Eqs. (%
\ref{N})-(15) in the case of%
\begin{equation}
0<k_{\rho }^{2}(w_{x}^{2}-w_{y}^{2})/4\ll 1.  \label{approx}
\end{equation}
This condition implies that either the soliton is wide in comparison with
the characteristic transverse-confinement width, $L_{z}$ (which may be
naturally expected from the quasi-2D solitons), i.e., $k_{\rho }\ll 1$, or
the profile is an almost symmetric one, with $\left\vert
w_{x}^{2}-w_{y}^{2}\right\vert \ll w_{x, y}^{2}$. Further analysis makes it
possible to expand, under condition (\ref{approx}) and to the first-order in $g_d$, the VA-predicted 2D norm
of the wave function as
\begin{eqnarray}
&&\hspace{-0.2in}N_{2}(\mu )=2\pi -\frac{g_{d}\,\pi ^{3/2}[1+3\cos (2\theta )]}{3(1+2\mu )^{2}%
\sqrt{\frac{-2}{\mu }-4}}\times   \label{N2} \\ \nonumber
&&\hspace{-0.2in}\left[ (4+2\mu )(4\mu -1)\sqrt{\frac{-1}{\mu }-2}+9\sqrt{2} \arctan%
\sqrt{\frac{-1}{2\mu }-1}\right],
\end{eqnarray}%
where $2\pi $ is the well-known VA prediction for the 2D norm of the Townes
solitons \cite{variation-1}, which is obviously valid in the limit of $%
g_{d}=0$, while the term $\sim g_{d}$ in Eq. (\ref{N2}) is a small
correction to it. The correction is a critically important one, as it lifts
the degeneracy of the Townes solitons, whose norm does not depend on $\mu $
\cite{Townes,Sulem,Fibich}, and thus makes it possible to check the VK\
criterion, which states that a necessary condition for the stability of any
soliton family supported by self-attractive nonlinearity is $dN_{2}/d\mu <0$
\cite{VK, vk1, vk2, vk3}. It originates from the condition that a soliton
which may be stable should realize a minimum of the energy for a given value
of the norm. Note also that condition $-1/2 < \mu < 0$, which is obviously
necessary for the validity of Eq. (\ref{N2}), definitely holds in the range
of $\mu $ given by Eq. (\ref{range}), dealt with in the present work.

Applying the VK criterion to the $N_{2}(\mu )$ dependence given by Eq. (\ref%
{N2}), we obtain
\begin{eqnarray}
&&\hspace{-0.2in}\frac{d\,N_{2}}{d\,\mu }=-\frac{\,g_{d}\pi ^{3/2}\left( 1-3\cos ^{2}\theta
\right) }{\mu (1+2\mu )^{3}\sqrt{\frac{-2}{\mu }-4}} \label{Vakhitov} \\ \nonumber
&&\hspace{-0.2in}\times \left[ 2\mu (4\mu -13)\sqrt{\frac{-1}{\mu }-2} +3\sqrt{2}(8\mu -1)\tan ^{-1}\sqrt{\frac{-1}{2\mu }-1}\right].
\end{eqnarray}%
It immediately follows
from Eq. (\ref{Vakhitov}) that the VK criterion holds, i.e., the
solitons \emph{may be stable} (in the framework of the VA), if the dipoles
are polarized under a sufficiently large angle $\theta $ with respect to the
normal direction, i.e., the polarization is relatively close to the in-plane
configuration (cf. Ref. \cite{anisotropic-soliton}):
\begin{equation}
\theta >\theta _{m}\equiv \cos ^{-1}(1/\sqrt{3})\approx 0.955~(\text{%
tantamount to }54.74^{\circ }).  \label{magic}
\end{equation}%
On the other hand, the solitons are predicted to be definitely unstable at $%
\theta <\theta _{m}$. The same critical (alias ``magic")
angle is known, e.g., in the theory of the nuclear magnetic resonance, when
a sample is spinning about a fixed axis \cite{magic-1, magic-2}. Note that,
in the framework of the approximation based on Eqs. (\ref{N2}) and (\ref%
{Vakhitov}), at $\theta =\theta _{m}$ the 2D norm of the solitons coincides
with that of the Townes solitons.

In the more general case, we have found the VA-predicted parameters $N_{2}$
and $w_{x,y}$ solving Eqs. (\ref{N})-(\ref{2}) numerically. In Fig.. \ref%
{fig6} we present the comparison of norm of $N_{2}$, as obtained from the
full numerical solution of Eq. (\ref{scaled-2D}) and its counterpart
predicted by the VA (solid and dashed curves, respectively). For the
reference, we also show the constant value, $N_{2}^{(T)}\approx 5.85$ for
the Townes solitons ($g_{d}=0$), and its above-mentioned VA-predicted
counterpart, $N_{2}^{(T)}=2\pi $ \cite{variation-1}. In particular, Fig. \ref%
{fig6}(a) features the same trends in the dependence $N_{2}(\theta )$ at
fixed $\mu $ as were identified, and qualitatively explained, above while
addressing Figs. \ref{fig4} and \ref{fig5}: in the case of the weak DDI, the
dependence is monotonous, while the strong DDI gives rise to a
well-pronounced maximum at point (\ref{m}).

In Fig. \ref{fig6}, we also depict the 2D norm $N_{2}$ as a function of the
chemical potential, $\mu $, for (b) weak and (c) strong DDI, i.e., $g_{d}=0.1
$ and $1.0$, respectively, for three fixed tilt angles, namely, $\theta =0$
(the dipoles polarized perpendicular to the pancake), $\theta = \theta_0$
[the special value  given by Eq. (\ref{m})], and $\theta =\pi /2$ (the in-plane polarization). In particular, it is seen that the slope of the $%
N_{2}\left( \mu \right) $ dependences, which determines the VK criterion, is
definitely positive, slightly or strongly positive (for small or large DDI strength), and slightly negative, for $\theta =0
$ (red curves), $\theta =\theta _{0}$ (magenta curves), and $\theta =\pi /2$
(blue curves), respectively. These conclusions, which pertain to the weak
and strong DDI alike, agree with the prediction of Eq. (\ref{Vakhitov}),
namely, $dN_{2}/d\mu < 0$ for $\theta >\theta _{m}$, and $dN_{2}/d\mu > 0$ for $\theta
<\theta _{m}$.

Lastly, Figs. \ref{fig6}(b,c) also show, as a reference for possible
experimental realization, the expected numbers of atoms in the solitons
created in the $^{52}$Cr condensate, transversely trapped under condition (%
\ref{Cr}).

\begin{figure}[t]
\includegraphics[width=8.4cm, height=6.0cm]{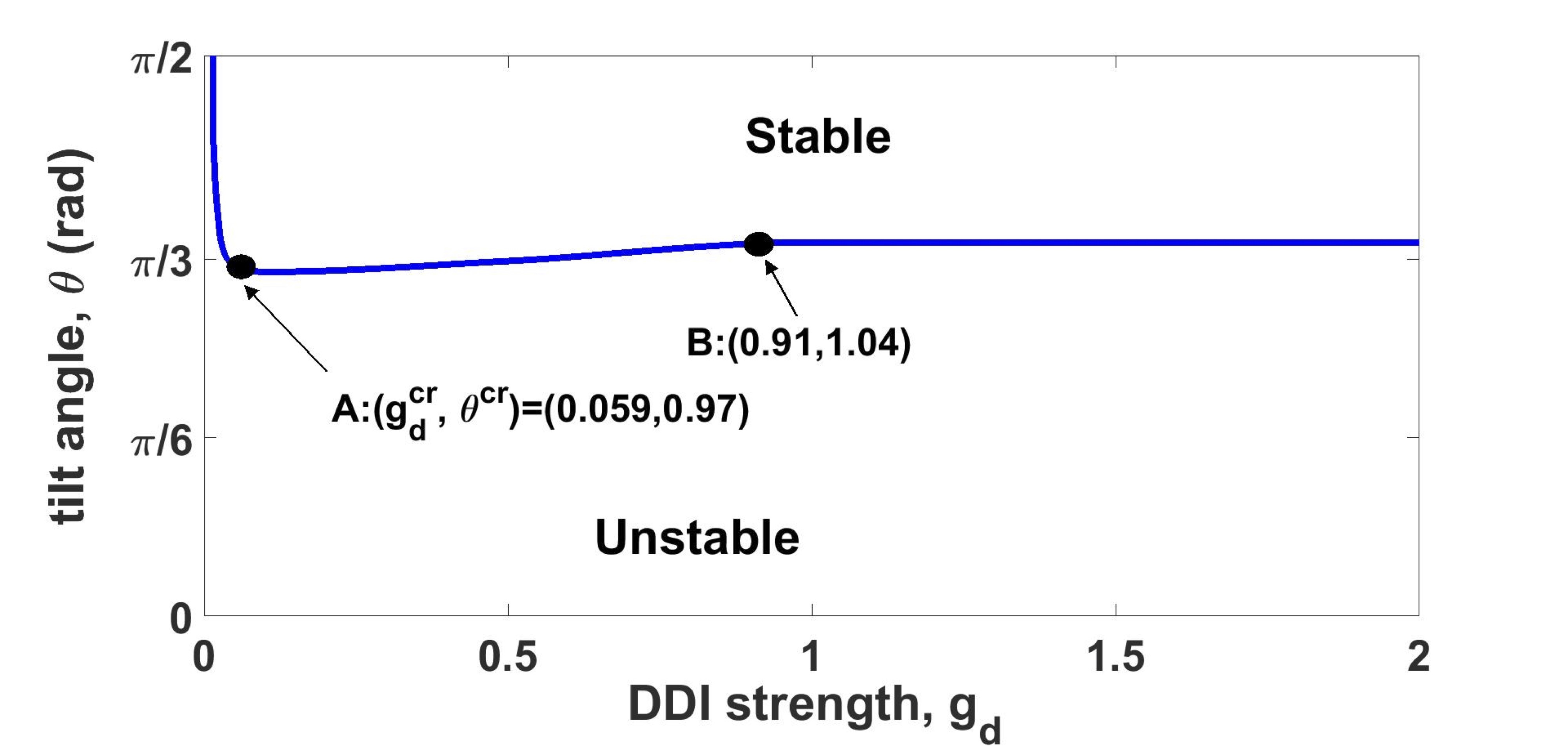}
\caption{The stability map for 2D solitons in the plane of the DDI strength,
$g_{d}$, and the tilt angle, $\protect\theta $, as produced by the solution
of the eigenvalue problem for small perturbations. The solitons are stable
at $\protect\theta ^{\text{cr}}<\protect\theta \leq \protect\pi /2$, where $%
\protect\theta ^{\text{cr}}(g_{d})$ is shown by the blue line. As above, the
strength of the contact interaction is fixed to be $g=-1$.  Points \textbf{A
}and\textbf{\ B }correspond, respectively, to the smallest and largest
values of $\protect\theta ^{\text{cr}}$, respectively (the definition of the
largest value excludes the narrow stripe of the quick decrease of $\protect%
\theta ^{\text{cr}}$ with the increase of $g_{d}$ from $0$ to point \textbf{A%
}). In fact, the difference between the largest and smallest values is
small. The nearly flat shape of the stability boundary roughly agrees with
the analytical prediction given by Eq. (\protect\ref{magic}).}
\label{fig7}
\end{figure}

\section{Stability of the 2D solitons}

As said above, stability is the critically important issue for 2D solitons,
as the usual cubic local self-attraction creates Townes solitons which are
subject to the subexponential instability against small perturbations \cite%
{BAMalomed,Sulem,Fibich}. Originally, the perturbations grow with time
algebraically, rather than exponentially, but eventually the solitons are
quickly destroyed. The subexponential instability implies that, in terms if
the above-mentioned VK criterion, the Townes solitons are, formally,
neutrally stable, having $dN_{2}/d\mu =0$ [see the flat black lines in Figs. %
\ref{fig6}(b,c)].

As said above, Eq. (\ref{Vakhitov}) and Figs. 6(b,c) demonstrate that the
addition of the DDI to the local self-attraction lifts the degeneracy (the
independence of the norm of the Townes solitons on the chemical potential).
The resulting sign of the slope, $dN_{2}/d\mu $, is the same for the
numerical solutions and their counterparts predicted by the variational
approximation. The sign is the same too for both the weak and strong DDI ($%
g_{d}=0.1$ and $g_{d}=1$). Equation (\ref{Vakhitov}) produces an important
prediction, that, with the increase of the title angle from $\theta =0$ to $%
\pi /2$, the slope $dN_{2}/d\mu $ changes from positive (unstable) to
negative (possibly stable) at the ``magic angle" given by
Eq. (\ref{magic}).

Because the VK criterion is only a necessary stability condition, and also
because Eq. (\ref{Vakhitov}) was derived in approximately, under condition Eq. (%
\ref{approx}), it is necessary to develop the consistent linear stability
analysis for our numerically generated soliton solutions. To this end, we
introduce a perturbed solution as

\begin{equation}
\psi (\rho ,t)=\{\psi _{0}(\rho )+\epsilon \left[ p(\rho )e^{-i\delta
t}+q(\rho )e^{i\delta ^{\ast }t}\right] \}e^{-i\mu t}.
\end{equation}%
Here, the asterisk stands for the complex-conjugate value, $\psi _{0}(\rho )$
is the unperturbed solution, $\epsilon $ is an infinitesimal perturbation
amplitude, while $p(\rho )$ and $q(\rho )$ are eigenmodes of the small
perturbation, with the respective eigenvalue $\delta $. The instability
occurs in the case when $\delta $ is not real. The unperturbed solution was
classified as a stable one if the numerically found instability growth rate,
$\left\vert \mathrm{Im}(\delta )\right\vert $, was smaller than $10^{-7}$.

Results of the stability analysis are summarized in Fig. \ref{fig7}, where
the stability map for the soliton solutions is displayed in the plane of the
DDI strength, $g_{d}$, and the tilt angle, $\theta $, the stability region
being
\begin{equation}
\theta ^{\mathrm{cr}}(g_{d})<\theta \leq \pi /2.  \label{cr}
\end{equation}%
This map is found to be\emph{\ the same}, up to the accuracy of the
numerically collected data, for the entire interval (\ref{range}) of values
of the chemical potential in which the derivation of the effective 2D
equation (\ref{scaled-2D}) is valid. This map shows that the originally
unstable Townes solitons, corresponding to $g_{d}=0$, quickly attains the
\emph{stability saturation}, i.e., expansion of the stability interval (\ref%
{cr}) to its limits, $\theta ^{\mathrm{cr}}\approx \theta _{m}<\theta \leq
\pi /2$, at very small values of $g_{d}$. At $g_{d}=g_{d}^{\text{cr}}\approx
0.059$, the stability boundary attains its minimum value, $\theta ^{\text{cr}%
}\approx 0.97$, as labeled by the point \textbf{A} in Fig. \ref{fig7}. With
the increase of $g_{d}$, the critical tilt slightly increases to $\theta
=1.04$ (tantamount to  $59.59^{\circ }$), as labeled by point \textbf{B},
which corresponds to $g_{d}=0.91$. Comparing these numerically exact results
with the analytical prediction given by Eq. (\ref{magic}), we conclude that
the relative error is limited to $8.2\%$, and, although the VA fails to
predict the dependence of $\theta ^{\text{cr}}$ on $g_{d}$, the actual
dependence is quite weak.

Lastly, it is relevant to stress that, setting $d N_2/d\mu = 0$ to identify 
the VK-predicted stability boundary, we obtain results, from the full numerical solution,
%$\theta ^{\text{cr}} = 0.98$ and $\pi/3 \approx 1.05$ 
for both weak and strong DDI,
with $g_d = 0.1$ and $1.0$, respectively, which
exactly coincide with the stability boundary identified 
above through the calculation of the linear-stability eigenvalues, 
i.e., $\theta ^{\text{cr}} = 0.97$ and $1.04$.

Before the conclusion, we discuss the possibility to stabilize dipolar BECs with quantum fluctuations. The stability boundary we reveal above is based on the mean-field theory. 
However, when the quantum fluctuations are taken into consideration, a repulsive, known as Lee-Huang-Yang (LHY), correction may stabilize an attractive Bose gas~\cite{65}.
Recently,  experimental observations on stable and ordered arrangement of droplets in an atomic dysprosium BEC illustrated the importance of LHY quantum fluctuations in stabilizing the system against collapse~\cite{66, 67}.
LHY corrections have be shown to stabilize droplets in unstable Bose-Bose mixtures \cite{68}, and self-bound filament-like droplets~\cite{69}. 
Relations on an arbitrary tilt angle to LHY corrections, and related stability of 2D solitons with DDI interaction deserve  further study.

\section{Conclusion}

For the dipolar BEC confined to the pancake geometry, we have investigated
the formation and stability of 2D soliton with the atomic magnetic moments
polarized in an arbitrary direction. Fixing the strength of contact
attractive interaction (which, by itself, would only create unstable Townes
solitons), we demonstrate, by means of the numerical methods and VA
(variational approximation), combined with the VK (Vakhitov-Kolokolov)
criterion, that the 2D solitons can be completely stabilized by the DDI
(dipole-dipole interaction) with relative strength $g_{d}$, which makes the
solitons anisotropic. Both the VK criterion and numerically exact
linear-stability analysis confirm that, there exists a
``magic angle" of the polarization tilt, $\theta ^{\text{cr}}$, such that the
2D solitons are stable at $\theta ^{\mathrm{cr}}(g_{d})<\theta \leq \pi /2$.
While the VA predicts $\theta ^{\mathrm{cr}}=\arccos (1/\sqrt{3})$ which
does not depend on $g_{d}$, the numerically exact results feature a weak
dependence of $\theta ^{\mathrm{cr}}$ on $g_{d}$, with the actual values of $%
\theta ^{\mathrm{cr}}$ being quite close to the VA prediction. We also
produce physical parameters for experiments in the condensate of $^{52}$Cr
atoms, which should make the creation of the stable 2D solitons possible.

\section*{ACKNOWLEDGMENTS}

This work was supported by the Ministry of Science and Technology of Taiwan
under Grant Nos. 105-2119-M-007-004.
The work of Y.L. was supported by Grant No. 11575063 from the National Natural Science Foundation of China.
The work of B.A.M. was supported, in part, by Grant No. 2015616 from the joint program in physics between the Binational (US-Israel) Science Foundation and National Science Foundation
(USA).

%%%%%%%%%%%%%%%%%%%%%%%%%%%%%%%%%%%%%%%%%%%%%%%%%%%%%%%%%%%%%%%%%%%%%%%%%%%%%%%%%%%%%%%%%%%%%%%%%%%%%%%%%%%%%%%%%%%%%%%%%%%%%%%%%%%%%%%%%%%%%%%%%%%%%%%%%%%%%%%%%%%%%%%%%%%%%%

\end{document}